\documentclass{article}

\usepackage{PRIMEarxiv}

\usepackage[utf8]{inputenc} 
\usepackage[T1]{fontenc}    
\usepackage{hyperref}       
\usepackage{url}            
\usepackage{booktabs}       
\usepackage{nicefrac}       
\usepackage{microtype}      
\usepackage{lipsum}
\usepackage{fancyhdr}       
\usepackage{graphicx}       
\graphicspath{{media/}}     

\usepackage{amsmath,amssymb,amsfonts}
\usepackage{algorithmic}
\usepackage{textcomp}
\usepackage{xcolor}
\usepackage[numbers,square]{natbib}
\usepackage{stfloats,framed}
\usepackage{geometry}

\title{Knowledge Distillation for Speech Denoising by Latent Representation Alignment with Cosine Distance
}

\author{
 Diep Luong \\
 Tampere University, Tampere, Finland \\
 Nokia Technologies, Espoo/Tampere, Finland\\
   \And
 Mikko Heikkinen \\
 Nokia Technologies, Espoo/Tampere, Finland \\
  \And
 Konstantinos Drossos \\
 Nokia Technologies, Espoo/Tampere, Finland \\
  \And
 Tuomas Virtanen \\
 Tampere University, Tampere, Finland \\
}

\begin{document}

\twocolumn[
\maketitle 
\begin{abstract}
Speech denoising is a generally adopted and impactful task, appearing in many common and everyday-life use cases. Although there are very powerful methods published, most of those are too complex for deployment in everyday and low-resources computational environments, like hand-held devices, intelligent glasses, hearing aids, etc. Knowledge distillation (KD) is a prominent way for alleviating this complexity mismatch and is based on the transferring/distilling of knowledge from a pre-trained complex model, the teacher, to another less complex one, the student. Existing KD methods for speech denoising are based on processes that potentially hamper the KD by bounding the learning of the student to the distribution, information ordering, and feature dimensionality learned by the teacher. In this paper, we present and assess a method that tries to treat this issue, by exploiting the well-known denoising-autoencoder framework, the linear inverted bottlenecks, and the properties of the cosine similarity. We use a public dataset and conduct repeated experiments with different mismatching scenarios between the teacher and the student, reporting the mean and standard deviation of the metrics of our method and another, state-of-the-art method that is used as a baseline. Our results show that with the proposed method, the student can perform better and can also retain greater mismatching conditions compared to the teacher. \vspace{0.15in}
\end{abstract}
]

\begin{figure*}[b!]
\begin{center}  
    \fbox{\strut\small Accepted for publication in the 158th Audio Engineering Society Convention proceedings as an Express Paper.}  
\end{center}
\end{figure*}

\section{Introduction}
Speech denoising is a widely explored task, focusing on recovering the speech signal from a noisy mixture~\cite{xu:2020:neurips}. Various approaches have been proposed, with deep learning-based methods established as the leading ones~\cite{huang:2015:interspeech,rethage:2018:icassp,kataria:2021:icassp}. However, state-of-the-art (SOTA) methods, like~\cite{hu:2020:interspeech}, typically feature intricate models with increased computational complexity. Such models need high computational resources that render them impractical for low-resource environments like hearing aids and smartphones, where computationally light-weight, low-latency, and/or real-time models are essential~\cite{choi:2021:icassp}. 

Various approaches have been proposed for reducing model complexity while retaining good performance to make them suitable for deployment in low-resource environments, with the three common approaches for the reduction of complexity for speech denoising to be: a) model pruning, b) quantization, and c) transfer learning-based methods. Model pruning-based approaches reduce the complexity of the speech denoising model by identifying and removing the learnable parameters that have the least contribution to the model output (i.e., pruning the model)~\cite{blalock:2020:mlsys,fang:2023:cvpr,he:2024:tpam,stamenovic:2021:neurips}. Quantization-based approaches aim at storing model parameters with fewer bits, thereby reducing memory and computation resources required when deploying the model in environments with low computational resources~\cite{yang:2019:cvpr,horowitz:2014:isscc,nicodemo:2020:eusipco}. Finally, transfer learning-based approaches aim at transferring learned information from an existing complex model to another less complex model. 

Knowledge distillation (KD) is a prominent transfer learning approach for reducing the complexity of speech denoising models by yielding a less complex model (the student) than an existing one (the teacher)~\cite{hinton:2015:nips}. Unlike pruning- or quantization-based approaches, KD seems to offer more flexibility in terms of model architecture and type, as different architectures and/or types of models (e.g., non-causal vs causal model) can be used as teacher and student~\cite{liu:2024:interspeech}. In a nutshell, KD involves the aligning of the information that is learned by the student with that learned by the teacher~\cite{hinton:2015:nips}. In speech denoising, the information alignment of the KD process is commonly performed either on the outputs of the two models or on the intermediate features. When the outputs of the two models are used, there is a minimization of a loss function on the outputs of the student and the teacher, trying to match the output of the student with the output of the teacher, when given the same input, for example as in response-based distillation~\cite{thakker:2022:interspeech}. When intermediate features are used, there is a minimization of a loss function on latent representations, trying to match the latent representation of the student with the one from the teacher, as in feature-based KD~\cite{nathoo:2024:icassp,han:2024:spl}.

However, when the latent representations are used, there is a need for accommodating for the dimensionality and ordering of information mismatching (e.g. CNN channels with same index between student and teacher do not necessarily hold same information). This issue is addressed (explicitly or implicitly) by using either an averaging process between the two different tensors, as in~\cite{nathoo:2024:icassp} where self-similarity matrices are used, or by using a learned alignment that, usually, is implemented by an attention mechanism, as in~\cite{park:2024:spl} where multi-head attention is employed. Both of these approaches require explicit selection of the dimensions to be taken into account (e.g. only time, only feature, or both) and bound the learning of the student in the learned latent distribution of the teacher~\cite{ham:2023:arxiv,park:2024:nsr}. In addition, the former approach has an explicit dependency on the dynamics of the batch (e.g. examples in the batch and size), while the second makes the implicit assumption, through the softmax activation in the attention mechanism, that there is a one-to-one (or close to one-to-one) mapping of the information between the two tensors used in the feature-based KD. Furthermore, recent findings~\cite{velickovic:2025:neurips} suggest that the use of the unregulated softmax (i.e., typical softmax) in the attention can be quite suboptimal when there is significant difference between the datasets used for the initial training of the teacher and the subsequent KD process. Finally, empirical evidence suggests that KD methods based on the above approaches, rely strongly on the numerical balance between the losses used for the alignment of the latent representations and the other losses used for the total KD-based training (e.g. response-based loss and/or typical supervised loss of the student). This means that the performance of the methods is sensitive to the weights used for the losses of the alignment process in KD.

In this paper we focus on the above mentioned issues and we propose a KD method that is using an implicit and learned ordering of the information but without having the above mentioned requirements. Our method is motivated by the denoising autoencoder (DAE) framework and previous work on KD~\cite{ham:2023:arxiv,park:2024:nsr} using cosine distance as KD loss, and is based on the linear and inverted bottlenecks and the feature-based approaches for KD, using a CNN-based linear bottleneck for dimensionality and information ordering matching of the latent representations. The cosine distance guides the student towards the learned, by the teacher, manifold of the clean speech, but without bounding the learning of the student on the learned distribution of the teacher. The linear bottleneck projects the latent representation that is used for KD to a lower dimensional space, without compromising the represented information~\cite{vincent:2010:jmlr,sandler:2018:cvpr}. 

We evaluate our method through an ablation study, using publicly available datasets, and we report the mean and standard deviation (STD) calculated over five repeated experiments. Our results show that our proposed method yields better and more robust (i.e. smaller STD) results compared to SOTA. The rest of the paper is as follows. Section~\ref{sec:method} presents necessary background and our proposed method and in Section~\ref{sec:evaluation} we present the evaluation setting and process. Section~\ref{sec:results} holds the results and related discussion and Section~\ref{sec:conclusions} concludes this paper. 
%
%
\section{Proposed method}\label{sec:method}
Our method takes as an input a pre-trained teacher model $\text{M}_{t}$, a randomly initialized student model $\text{M}_{s}$, and a dataset $\mathbb{D}=\{(\tilde{\mathbf{X}}^{n}, \mathbf{X}^{n})\}_{n=1}^{N_{i}}$, of $N_{i}$ noisy ($\tilde{\mathbf{X}}$) and corresponding clean speech ($\mathbf{X}$) examples, and outputs the optimized student model $\text{M}_{s}^{\star}$ which extracts and exploits information that is aligned with the one from the pre-trained teacher model. For this process, our method minimizes the joint loss $\mathcal{L}_{tot} = \mathcal{L}_{kd} + \mathcal{L}_{out}$, consisting of one KD loss on the latent representations, $\mathcal{L}_{kd}$, and one supervised loss, $\mathcal{L}_{out}$, between the denoised signal from $\text{M}_{s}$ and $\mathbf{X}$. $\mathcal{L}_{kd}$ is implemented by cosine distance while $\mathcal{L}_{out}$ is a typical speech denoising loss, which, in this study, is a source separation based loss like SI-SNR~\cite{luo:2018:icassp,roux:2019:icassp}. 
\subsection{Teacher and student models}
Our method assumes that both $\text{M}_{t}$ and $\text{M}_{s}$ are based on the DAE architecture, i.e. both consist of two parameterized functions; an encoder $\text{E}_{*}$, and a decoder, $\text{D}_{*}$\footnote{In the rest of the text the symbol $*$ will be used as a wildcard for indicating teacher and/or student related symbols and variables.}. Additionally, both $M_{*}$ take as input the magnitude of the time-frequency (TF) representation of $\tilde{\mathbf{X}}$, denoted as $\tilde{\mathbf{Y}} = Z(\tilde{\mathbf{X}})$. Here, $\tilde{\mathbf{Y}} \in \mathbb{R}^{T \times F}$ is the magnitude of the TF representation of $\tilde{\mathbf{X}}$, consisting of $T$ TF vectors each of length $F$. The function $\text{Z}$ is responsible for extracting the magnitude of the TF representation, using methods such as FFT or STFT.


Given the DAE architecture, the encoder consists of $N_{\text{E},*}$ blocks, while the decoder is comprised of $N_{\text{D},*}$ blocks. It is important to note that the number of blocks in the encoder and decoder are equal, i.e. $N_{\text{E},*}=N_{\text{D},*}=N_*$. Within the encoder, there are $N_*$ sequential processes $P_{\text{E}, *}^{n}$, with $n = 1, 2, \ldots, N_*$, as
\begin{equation}
    \text{E}_{*} = P_{\text{E}, *}^{N_*}\circ P_{\text{E}, *}^{N_*-1} \circ \ldots \circ  P_{\text{E}, *}^{1}.
\end{equation}
\noindent
$\text{E}_{*}$ takes as an input $\tilde{\mathbf{Y}}$ and outputs its latent representation $\mathbf{H}_{*}$ as
\begin{align}
    \mathbf{H}_{\text{E}, *} &= \text{E}_{*}(\tilde{\mathbf{Y}}), \text{ where}\\
    \mathbf{H}_{\text{E}, *} &= P_{\text{E}, *}^{N_*}(\mathbf{H}_{\text{E}, }^{N_*-1}),\\
    \mathbf{H}_{\text{E}, *}^{n} &= P_{\text{E}, *}^{n}(\mathbf{H}_{\text{E}, }^{n-1}),\\
    \mathbf{H}_{\text{E}, *}^{0} &= \tilde{\mathbf{Y}}, 
\end{align}
\noindent
$\mathbf{H}_{\text{E}, *} \in \mathbb{R}^{C_{\text{E}, *}\times H_{\text{E}, *}\times W_{\text{E}, *}}$ is the output latent representation of $\text{E}_{*}$ where $C_{\text{E}, *}\geq1$ $H_{\text{E}, *} < T$, and $W_{\text{E}, *} < F$. 

$\text{D}_{*}$, similarly to $\text{E}_{*}$, consists of $N_*$ concatenated processes $P_{\text{D}, *}^{n}$, with $n = 1, 2, \ldots, N_*$, as
\begin{equation}
    \text{D}_{*} = P_{\text{D}, *}^{N_*}\circ P_{\text{D}, *}^{N_*} \circ \ldots \circ  P_{\text{D}, *}^{1}.
\end{equation}
\noindent
$\mathbf{H}_{\text{E}, *}$ is given as an input to $\text{D}_{*}$ as
\begin{align}
    \hat{\mathbf{Y}}_{*} &= \text{D}_{*}(\mathbf{H}_{\text{E}, *}), \text{ where}\\
    \hat{\mathbf{Y}}_{*} &= P_{\text{D}, *}^{N_*}(\mathbf{H}_{\text{D}, *}^{N_*-1}, \mathbf{H}_{\text{E}, * }^{1}),\\
    \mathbf{H}_{\text{D}, *}^{n} &= P_{\text{D}, *}^{n}(\mathbf{H}_{\text{D}, *}^{n-1}, \mathbf{H}_{\text{E}, *}^{N_*-(n-1)})\text{ for } n \geq 2, \\
    \mathbf{H}_{\text{D}, *}^{1} &= P_{\text{D}, *}^{1}(\mathbf{H}_{\text{D},*}^{0}),\\
    \mathbf{H}_{\text{E}, *} &= \mathbf{H}_{\text{D}, *}^{0}.
\end{align}
\subsection{Dimensionality and information ordering matching}
Our method uses the $\mathbf{H}_{\text{E}, *}$ for the KD process. Though, there is the flexibility for dimensionality mismatching between $\mathbf{H}_{\text{E}, t}$ and $\mathbf{H}_{\text{E}, s}$, i.e. at least one of $C_{\text{E}, t} \neq C_{\text{E}, s}$, $H_{\text{E}, t} \neq H_{\text{E}, s}$, and $W_{\text{E}, t} \neq W_{\text{E}, s}$ is true. But, even if there is no dimensionality mismatching, given that $\text{E}_{t}$ and $\text{E}_{s}$ have different parameters and (given the scope of this work) most likely considerable difference in the number of learned parameters, then it is almost certain that there is no consistent ordering of the information between $\mathbf{H}_{\text{E}, t}$ and $\mathbf{H}_{\text{E}, s}$, e.g. $\mathbf{H}_{\text{E}, t}[c, :] \neq \mathbf{H}_{\text{E}, s}[c, :]$. 

To solve the above, our method uses a linear bottleneck, $\text{B}_{t}$, that maps the dimensionality and/or re-orders the information of $\mathbf{H}_{\text{E}, t}$ to match that of $\mathbf{H}_{\text{E}, s}$. This choice is motivated by the usage of convolutional linear bottleneck in the MobileNetv2 architecture~\cite{sandler:2018:cvpr}, where a linear bottleneck was used to reduce the dimensionality of the feature map but without (or without considerable) loss of information~\cite{sandler:2018:cvpr,han:2017:cvpr}. Therefore, our method adopts the linear bottleneck to, on one hand, map the dimensionality of the $\mathbf{H}_{\text{E}, t}$ to the dimensionality of the $\mathbf{H}_{\text{E}, s}$ and, on the other hand, to allow for learnable re-ordering of information between the dimensions of the $\mathbf{H}_{\text{E}, *}$. Specifically, our method uses $\text{B}_{t}$ as
\begin{equation}
    \mathbf{H}_{\text{B}, t} = \text{B}_{t}(\mathbf{H}_{\text{E}, t}),
\end{equation}
\noindent
where $\mathbf{H}_{\text{B}, t} \in \mathbb{R}^{C_{\text{E}, s}\times H_{\text{E}, s}\times W_{\text{E}, s}}$.

\subsection{Losses and student optimization}
\begin{figure}
    \centering
    \includegraphics[width=0.8\linewidth]{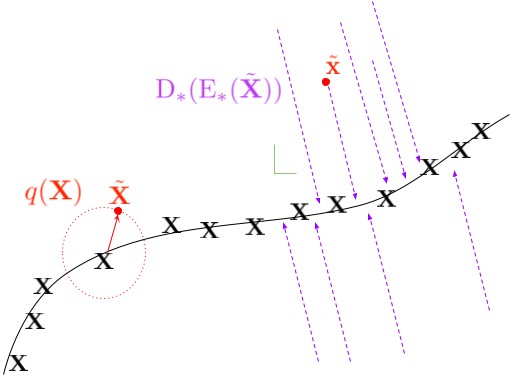}
    \caption{Illustration of the DAE process, where $q(\cdot)$ is a corruption process that yields $\tilde{\mathbf{X}}$. With red color are indicated the noisy/corrupted data ($\tilde{\mathbf{X}}$), with black the clean ones ($\mathbf{X}$), with purple is the mapping that is learned from the DAE, and the green axis are for visual convenience. Image is after~\cite{vincent:2010:jmlr}.}
    \label{fig:dae-manifold}
\end{figure}

Motivated by the DAE framework, our method uses the cosine distance as the KD loss. As shown in~\cite{vincent:2010:jmlr} and is illustrated in Figure~\ref{fig:dae-manifold}, a DAE tries to map a noisy sample to the manifold of the source. In the scope of this work, the pre-trained $\text{M}_{t}$ is already optimized to implement the mapping of noisy speech samples to the manifold of the clean speech, i.e. perform denoising of the noisy input $\tilde{\mathbf{X}}$. Although the latent features learned by $\text{M}_{s}$ are not necessarily in the same numerical range as the ones learned by $\text{M}_{t}$, especially if these features are observed early in the model (e.g. at the end of $\text{E}_{*}$), the two need to be oriented towards the same manifold, i.e., that of the clean speech, since the supervised training target of both $\text{M}_{t}$ and $\text{M}_{s}$ is the same. Additionally, given the learning/capacity gap between $\text{E}_{t}$ and $\text{E}_{s}$, the distribution of the latent features that are learned by $\text{E}_{t}$ will (most likely if not certainly) are different from that which $\text{E}_{s}$ will learn. Using a divergence metric (e.g. KL) or an $\text{L}_{p}$ based metric as the KD loss, will bound the learning of the $\text{E}_{s}$~\cite{ham:2023:arxiv,park:2024:nsr}. 

The above observations motivated our method to adopt the cosine distance ($\text{cos}_{dist}$) as a KD loss. Since $\text{cos}_{dist}$ is scale-invariant, the KD process is expected to align the angle of the latent features between $\text{E}_{t}$ and $\text{E}_{s}$, without trying to match either the numerical ranges or the distribution of the latent features. This is most likely to yield more learning and aligning capacity at the KD process~\cite{ham:2023:arxiv,park:2024:nsr}. Our method uses $\text{cos}_{dist}$ as
\begin{align}
    \mathcal{L}_{kd} =&\text{ }\text{cos}_{dist}(\mathbf{H}_{\text{B}, t}, \mathbf{H}_{\text{E}, s}),\text{ where}\\
    \text{cos}_{dist}(\mathbf{H}_{\text{B}, t}, \mathbf{H}_{\text{E}, s}) =&\text{ }1 - \frac{\langle\mathbf{H}_{\text{B}, t}, \mathbf{H}_{\text{E}, s}\rangle}{||\mathbf{H}_{\text{B}, t}||_{2} \cdot ||\mathbf{H}_{\text{E}, s}||_{2}},
\end{align}
\noindent
$\langle\cdot,\cdot\rangle$ is the dot product, and $||\cdot||_{2}$ is the Euclidean norm.

The learning of the $\text{M}_{s}$ is guided in a supervised way by employing a typical function for source separation tasks, $\text{S}_{s}$, (like mean squared error or SI-SNR) over the predicted output and the ground truth as
\begin{equation}
    \mathcal{L}_{out} = \text{S}_{s}(\hat{\mathbf{X}}, \mathbf{X}),\text{ where}
\end{equation}
\noindent

$\hat{\mathbf{X}} = \tilde{\mathbf{X}} \odot \text{M}_{s}(\tilde{\mathbf{X}})$ is the denoised signal, and $\mathbf{X}$ is the ground truth.

Finally, $\text{M}_{s}$ is optimized by the minimization of the joint loss $\mathcal{L}_{tot}$
\begin{equation}
    \mathcal{L}_{tot} = \lambda_{kd}\mathcal{L}_{kd} + \lambda_{out}\mathcal{L}_{out},\text{ where}\label{eq:joint-loss}
\end{equation}
\noindent
$\lambda_{kd}$ and $\lambda_{out}$ are scaling factors for the corresponding losses. 
%
%
\section{Evaluation}
\label{sec:evaluation}
We evaluate our method using a popular and widely adopted DAE architecture for speech denoising, a publicly available dataset, and well established metrics in speech denoising, and for all hyper-parameters we use as much as possible typical default options. Using the employed dataset we train a large teacher model, $\text{M}_{t}$, following typical set-up of hyper-parameters. Then we freeze the weights of $\text{M}_{t}$, employ the same dataset and the joint loss in Eq.~\ref{eq:joint-loss}, and optimize a randomly initialized small model, $\text{M}_{s}$. Both of the models, $\text{M}_{*}$, are based on the same architecture. To evaluate the performance of the choices for the KD process, we use three settings regarding the discrepancies between $C_{E,*}$, $H_{E,*}$, and $W_{E,*}$. Finally, we repeat each KD process 5 times and report the mean and standard deviation. 
\subsection{Dataset and data pre- and post-processing}
For the pre-training of the $\text{M}_{t}$ and the KD process, we use the publicly available DNS-Challenge dataset, specifically the v5 speakerphone set~\cite{dubey:2023:icassp} (referred to as development training).
We resample all files to 16kHz sampling rate and we manually split the speaker and noise files of the development training set to training, validation, and testing splits, with a ratio of (as close as possible to) 60/20/20\%, respectively, and ensuring that the splitting ratio is kept throughout the sub-sets of the DNS-Challenge (e.g. read speech, emotional speech, etc.). Then, for each split we segment all files to segments of two second duration, discarding segments with no activity or shorter than two seconds.

We mix clean speech and noise files by uniformly sampling integer SNR values from the range of $[-5, 20]$ dB SNR. All resulting noisy files, $\tilde{\mathbf{X}}$, are scaled to be in the limit of $[-1, 1]$ amplitude, also proportionally scaling the corresponding clean speech signal, $\mathbf{X}$. For both the pre-training and the KD process, the datasets $\mathbb{D}$ used for training and validation are formed on-the-fly, meaning that in each epoch the noise files were randomly selected (and, if needed, oversampled) from the pool of noise files. Though, the testing split was created once, to keep a consistent point of reference. From each two-second-long signal, we extract $T=126$ STFT vectors of $F=256$ elements, using 512 FFT points with 50\% overlap. We calculate the magnitude spectrum and phase from the STFT and use the magnitude as an input to $\text{M}_{*}$. The inverse STFT is performed using the predicted magnitude spectrum from the speech denoising process and the phase of the noisy input. 

\subsection{DAE architecture and linear bottleneck implementation}
For $\text{M}_{*}$ we use the Unet architecture~\cite{unet}, as it is a prominent and widely used architecture in speech denoising~\cite{riahi:2023:ccece,walczyna:2024:aiml,choi:2021:icassp,giri:2019:waspaa,stoller:2018:ismir}. To be able to evaluate our KD method under different dimensionality mismatching conditions at the $\mathbf{H}_{\text{E}, *}$, we use two different $\text{M}_{t}$, $\text{M}_{t1}$ and $\text{M}_{t2}$, and two different $\text{M}_{s}$, $\text{M}_{s1}$ and $\text{M}_{s2}$. $\text{E}_{t1}$ consists of $N_{\text{E},t1}=6$ CNN blocks and $\text{E}_{t2}$ of $N_{\text{E},t2}=7$. Each block has a 2D convolutional neural network (CNN) implementing a strided convolution, a normalization process (instance norm), and a nonlinearity (leaky rectified linear unit, ReLU, with default slope). Following typical practices, in $\text{E}_{t1}$ each CNN block halves the number of columns of its input tensor, while doubling the number of channels, while in $\text{E}_{t2}$ the halving of columns and the doubling of channels happens every two CNN blocks. Preserving the number of rows (i.e. corresponding to time dimension) is adopted according to most practices regarding the teacher model in KD and for speech denoising. In both $\text{E}_{t1}$ and $\text{E}_{t2}$ a square $5x5$ kernel with a stride of $1x1$ was used (for not affecting the dimensions) or $1x2$ if the columns were halved, and appropriate padding. The above yield $\{C_{\text{E}, t1}, H_{\text{E}, t1}, W_{\text{E}, t1}\} = \{128, 126, 5\}$ and $\{C_{\text{E}, t2}, H_{\text{E}, t2}, W_{\text{E}, t2}\} = \{128, 126, 17\}$.

\begin{figure}
    \centering
    \includegraphics[width=0.9\linewidth]{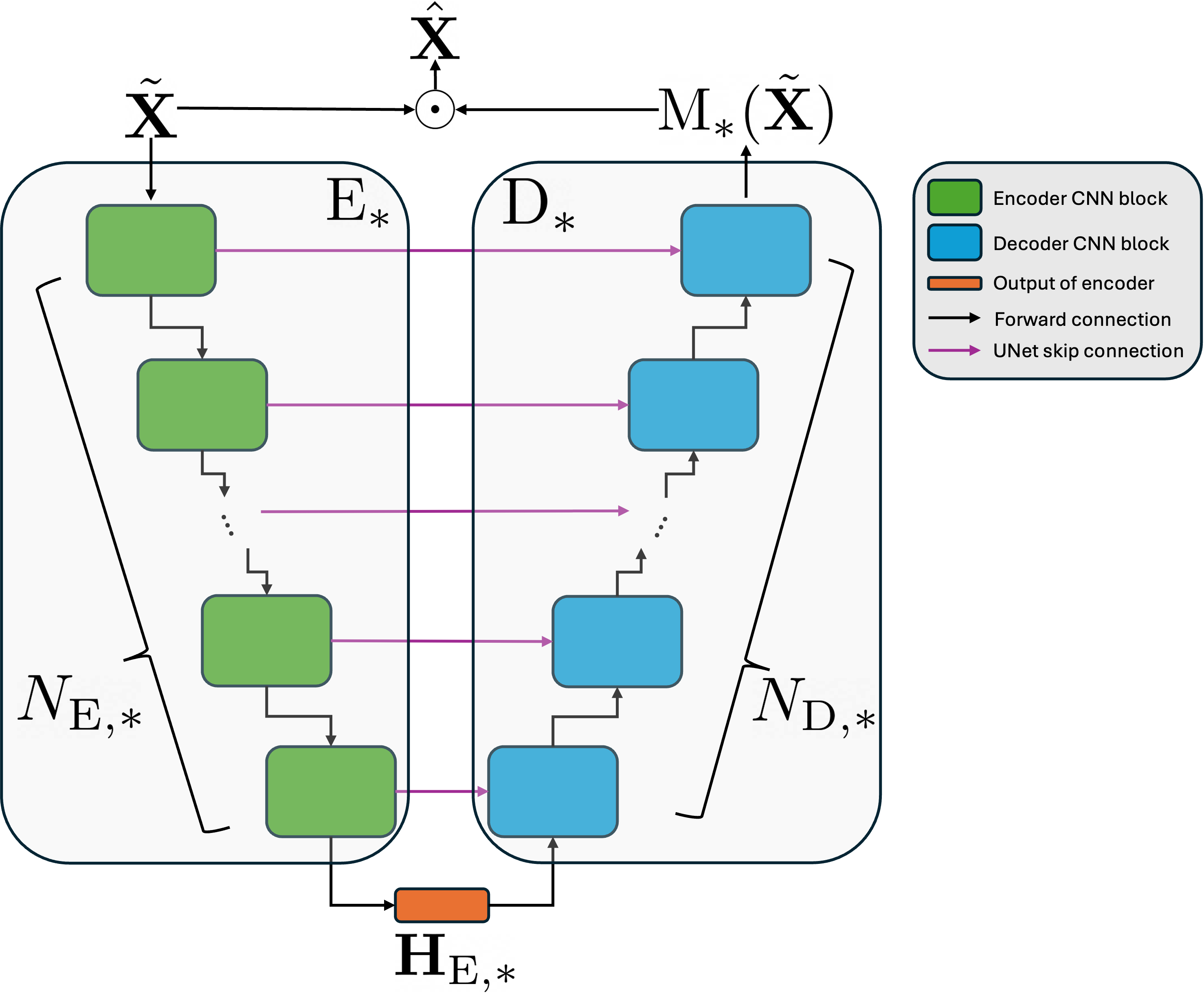}
    \caption{Illustration of the employed UNet architecture for the utilized DAEs. With $\odot$ is the Hadamard product}
    \label{fig:unet-arch}
\end{figure}

Followingly, $\text{D}_{t1}$ and $\text{D}_{t2}$ are symmetrical to the corresponding encoders, consisting of $N_{\text{D},t1}=6$ and $N_{\text{D},t2}=7$ transposed CNN blocks, respectively. Each block has a transposed CNN, a normalization process, and a non-linearity, similarly to $\text{E}_{t}$ and corresponding kernel, stride, and padding. The skip connections between $\text{E}_{t}$ and $\text{D}_{t}$ are implemented according to standard practices, using the output of each CNN block from the encoder, apart from the last one. Also, the last transposed CNN block has no normalization process and has sigmoid as non-linearity for both $\text{D}_{t1}$ and $\text{D}_{t2}$. The number of trainable parameters in $\text{M}_{t1}$ and $\text{M}_{t2}$ are 1.35M and 2.04M. Teacher models $\text{M}_{t1}$ and $\text{M}_{t2}$ has respectively 2996.20 MOps/sample and 10939.91 MOps/sample (i.e. the number of operations required to process a single sample). An illustration of the employed architecture is in Figure~\ref{fig:unet-arch}. 

$\text{E}_{s1}$ and $\text{E}_{s2}$ have $N_{\text{E},s1}=N_{\text{E},s2}=6$ cascaded CNN blocks, similarly to $\text{M}_{t}$, but with a square $3x3$ kernel. In $\text{E}_{s1}$ the stride is $1x2$, so the time-corresponding dimension is not altered, while in $\text{E}_{s2}$ a stride of $2x2$ is used and the time-corresponding dimensions is reduced as well. In both $\text{E}_{s1}$ and $\text{E}_{s2}$, the frequency-corresponding dimension (i.e. columns) are halved with every CNN block. The above yield $\{C_{\text{E}, s1}, H_{\text{E}, s1}, W_{\text{E}, s1}\} = \{32, 126, 5\}$ and $\{C_{\text{E}, s2}, H_{\text{E}, s2}, W_{\text{E}, s2}\} = \{32, 2, 5\}$. $\text{D}_{s1}$ and $\text{D}_{s2}$ are implemented symmetrically to $\text{E}_{s1}$ and $\text{E}_{s2}$ (i.e., same kernel size, stride, and padding) and according to $\text{D}_{t1}$ and $\text{D}_{t2}$ (i.e. same configurations for normalization and non-linearities). 
Both $\text{M}_{s1}$ and $\text{M}_{s2}$ have 37k trainable parameters, but differ in computational complexity, with $\text{M}_{s1}$ requiring 65.01 MOps/sample and $\text{M}_{s2}$ requiring 7.10 MOps/sample.

The linear bottleneck is implemented by a CNN block of concatenated CNNs, without any non-linearities or normalization processes in between as proposed in~\cite{sandler:2018:cvpr} for the inverted linear bottleneck module. The CNNs have a kernel of $1x1$ size, with unit stride and no padding. Effectively, this is an affine transform over the channel dimension of the input tensor. For example, in the case where the $C_{\text{E}, *}$ and $H_{\text{E}, *}$ needed to match between $\mathbf{H}_{\text{E}, t}$ and $\mathbf{H}_{\text{E}, s}$, the linear bottleneck consists of two CNNs. The first one takes as input $\mathbf{H}_{\text{E}, t}$ and operates along the $C$ dimension and the next one takes as input the output of the first and operates along the $H$ dimension. To test all cases of dimensionality mismatching, we used three different set-ups of bottlenecks; a) with one CNN, matching only $C_{*}$, b) with two CNNs, matching $C_{*}$ and then $H_{*}$, and c) with three CNNs, matching $C_{*}$, $H_{*}$, and $W_{*}$. 

\subsection{Training, knowledge distillation process, and metrics}
The pre-training of the $\text{M}_{t1}$ and $\text{M}_{t2}$ implemented by minimizing the SI-SNR loss
\begin{align}
    &\mathcal{L}_{pre} =\text{SI-SNR}(\mathbf{X}, \tilde{\mathbf{X}} \odot \text{M}_{t*}(\tilde{\mathbf{X}})),\text{ where}\\
    &\text{SI-SNR}(x, \hat{x}) =10\log_{10}(\frac{||\frac{\hat{x}^{T}x}{||x||^{2}}x||^{2}}{||\frac{\hat{x}^{T}x}{||x||^{2}}x-\hat{x}||^{2}}),
\end{align}
\noindent
$x$ is a target signal, and $\hat{x}$ is a prediction of $x$. Then, we implement the KD process by minimizing the loss in Eq.\ref{eq:joint-loss} and using the pre-trained teacher model. Both training of the teacher models and KD processes are implemented using Adam optimizer~\cite{adam} with default hyper-parameters, a batch size of 32, and adopting the early stopping policy on the loss of the validation split.

\begin{figure}
    \centering
    \includegraphics[width=0.9\linewidth]{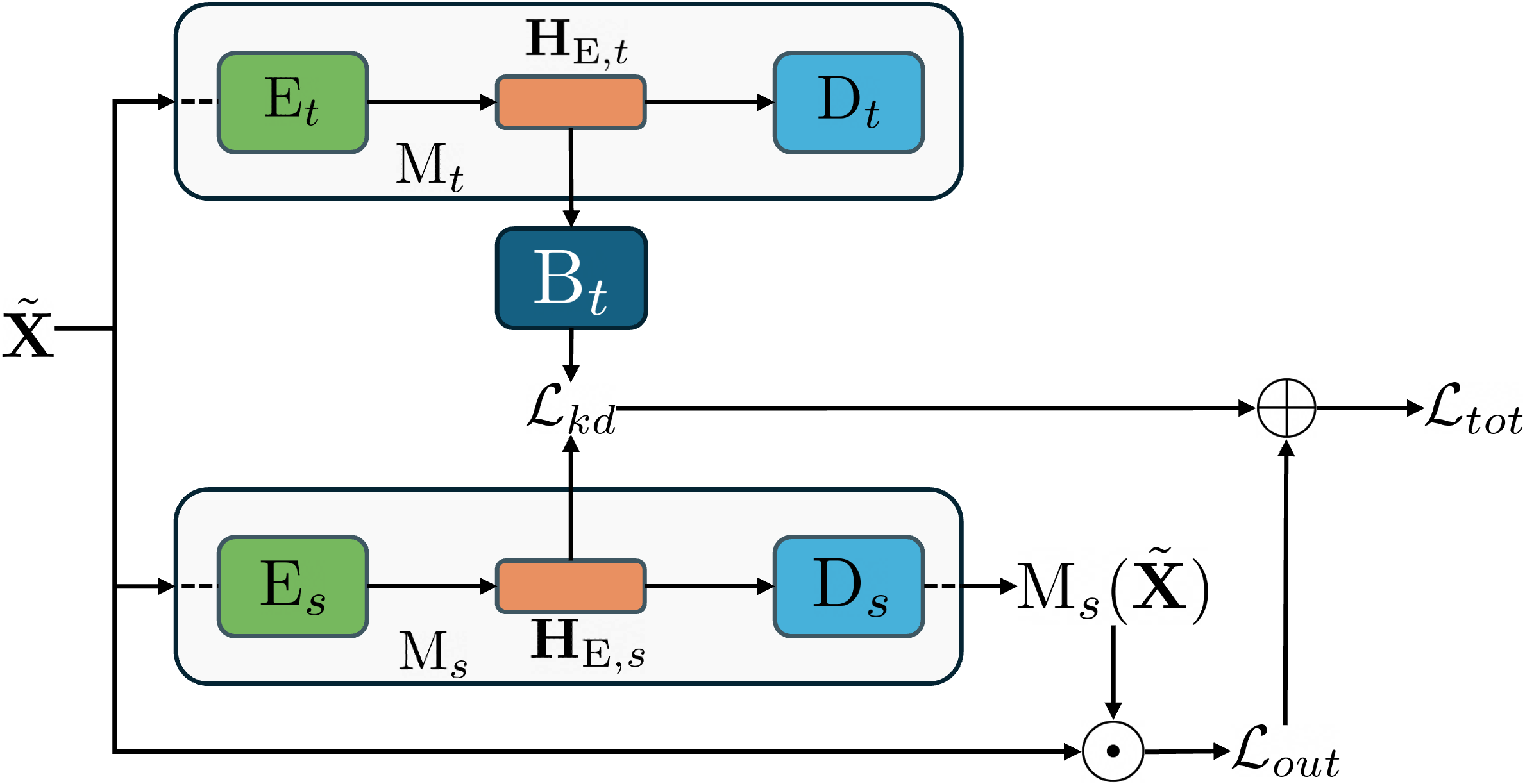}
    \caption{Illustration of the KD process, where $\oplus$ is simple summation.}
    \label{fig:kd-process}
\end{figure}

We implemented three KD processes; i) from $\text{M}_{t1}$ to $\text{M}_{s1}$ to evaluate the mismatching of the channel dimension (i.e. matching $C_{*}$) and/or the information rearrangement in the other dimensions, ii) from $\text{M}_{t1}$ to $\text{M}_{s2}$ to evaluate the mismatching of the channel dimension and the time dimension (i.e. matching $C_{*}$ and $H_{*}$) and/or the information rearrangement in the frequency dimension, and iii) from $\text{M}_{t2}$ to $\text{M}_{s2}$ to evaluate the mismatching of the channel, time, and frequency dimensions (i.e. matching $C_{*}$, $H_{*}$, and $W_{*}$). An illustration of the KD process is in Figure~\ref{fig:kd-process}. For evaluating our method without any optimization regarding the numerical balance of the losses, we opted for $\lambda_{kd}=\lambda_{out}=1$.

For the evaluation of the performance we employed the objective, intrusive metrics SDR, SI-SDR, wide-band PESQ, and STOI on the utilized testing split of the development training dataset.
While we pre-trained each teacher model just once, we implemented each of the three KD scenarios five times and we report the mean and STD of the above metrics. 

\subsection{Baselines}
We compare our method with SOTA KD methods for speech denoising~\cite{nathoo:2024:icassp,han:2024:spl}. For having a vanilla-based and fair comparison, we re-implemented these methods given the employed models in this work. Although we tried to re-implement the method in~\cite{nathoo:2024:icassp}, we could not reproduce any improvements on the student. Our empirical observation showed that there was a great fluctuation of performance in different, repeated experiments and that there is a great dependency on the numerical balance of the utilized losses. Though, we were able to do so with the method in~\cite{han:2024:spl}, which we use to compare our method with and we signify it as ATKL. 
%
%
\begin{table*}[!ht]
\caption{Mean/STD of evaluation metrics for our method and the baseline. ``$\text{A }\mapsto \text{ B}$'' shows KD process, from A to B. (C) means that there was a dimensionality mismatch handling only for the channel dimension and, similarly, (C, H) and (C, H, W) show the same for channel, time, and frequency dimensions, respectively. With ATKL is the baseline method, KD process results with ``-'' indicate that the KD process could not be implemented. Best results for each student model are indicated with bold.}
\centering
\label{tab:results-main}
\begin{tabular}{lllll}
\multicolumn{1}{c}{Model} & \multicolumn{1}{c}{SDR} & \multicolumn{1}{c}{SI-SDR} & \multicolumn{1}{c}{WB PESQ} & \multicolumn{1}{c}{STOI} \\
\hline
\multicolumn{1}{c}{T1} & 18.6882/0.0000 & 17.7157/0.0000 & 2.5323/0.0000 & 0.9178/0.0000\\
\multicolumn{1}{c}{T2} & 18.9415/0.0000 & 17.9524/0.0000 & 2.5419/0.0000 & 0.9194/0.0000\\
\multicolumn{1}{c}{S1} & 16.1381/0.0828 & 15.1891/0.0581 & 2.0239/0.0147 & 0.8837/0.0013\\
\multicolumn{1}{c}{S2} & 13.9856/0.0719 & 13.1243/0.0808 & 1.7537/0.0181 & 0.8502/0.0006\\
\hline\hline
\multicolumn{5}{c}{KD process with student S1}\\
\hline\hline
T1 $\mapsto$ S1 ATKL & 16.1308/0.0961 & 15.1848/0.0506 & \textbf{2.0365/0.0181} & 0.8839/0.0009\\
T1 $\mapsto$ S1 (C) & 16.0945/0.1102 & 15.2156/0.0522 & 2.0212/0.0143 & 0.8843/0.0010\\
T1 $\mapsto$ S1 (C, H) & 16.2008/0.1203 & \textbf{15.2445/0.0462} & 2.0246/0.0175 & \textbf{0.8845/0.0011}\\
T1 $\mapsto$ S1 (C, H, W) & \textbf{16.2162/0.0500} & 15.2059/0.0305 & 2.0176/0.0120 & 0.8837/0.0004\\
\hline\hline
\multicolumn{5}{c}{KD process with student S2}\\
\hline\hline
T1 $\mapsto$ S2 ATKL & 13.9380/0.0955 & 13.0617/0.0821 & 1.7393/0.0080 & 0.8478/0.0012\\
T2 $\mapsto$ S2 ATKL & \multicolumn{1}{c}{-} & \multicolumn{1}{c}{-} & \multicolumn{1}{c}{-} & \multicolumn{1}{c}{-}\\
T1 $\mapsto$ S2 (C, H) & 14.0517/0.1166 & \textbf{13.1663/0.0542} & \textbf{1.7594/0.0157} & \textbf{0.8515/0.0018}\\
T1 $\mapsto$ S2 (C, H, W) & \textbf{14.0856/0.0400} & 13.1415/0.0917 & 1.7429/0.0107 & 0.8496/0.0011 \\
T2 $\mapsto$ S2 (C, H, W) & 14.0277/0.0579 & 13.1606/0.0757 & 1.7528/0.0131 & 0.8505/0.0013
\end{tabular}%
\end{table*}

\section{Results and discussion}
\label{sec:results}
In Table~\ref{tab:results-main} are the obtained results on the testing split of the development training set. We report the mean and STD of the employed evaluation metrics, i.e. SDR, SI-SDR, wide band PESQ (WB-PESQ), and STOI, for each approach. 

As illustrated in Table~\ref{tab:results-main}, the proposed KD method demonstrates better performance compared to the ATKL baseline in all but one scenario. This indicates a consistent improvement across various settings, highlighting the robustness and effectiveness of the proposed approach. A particularly observation is 
the $T2 \mapsto S2$ configuration, a setting where ATKL fails to effectively adapt to the dimensionality mismatch. This suggests that the proposed method has greater flexibility and adaptability in transferring knowledge between teacher and student networks, even in challenging scenarios where ATKL may struggle.

Of all the cases of dimensionality matching, the benefits of the proposed method are more pronounced in the $(C, H)$ cases. Unlike averaging-based methods (e.g., ATKL), which may oversimplify the transfer process and potentially lose valuable information, the proposed method appears to retain and convey more nuanced and detailed knowledge from the teacher to the student network. This leads to better alignment and improved learning outcomes for the student network.

Additionally, from Table \ref{tab:results-main} can be observed that there are small numerical differences between the metrics of all cases. It should be noted that, as mentioned in Sections~\ref{sec:method} and~\ref{sec:evaluation}, the evaluation was kept as vanilla as possible. This was done in order to study and exhibit the robustness of the proposed method (i.e., the proposed method should work better even when nothing is specifically tuned for the use case), and to reveal the performance of the proposed method and the utilized baseline(s) in "default" scenarios. Given these, the observed increases are in accordance with existing work on speech denoising KD, where, quite often, increases in the range of 0.1 to 1 dB are observed for the student~\cite{nathoo:2024:icassp}. On the same note, the performance improvement of the student networks facilitated by KD, whether through ATKL or our proposed method, is marginal. This observation warrants a deeper investigation into the underlying reasons for such limited enhancements. One plausible explanation for this phenomenon is that the student network, when trained solely through supervised learning, might already reach its optimum in speech denoising. Consequently, the additional knowledge distilled from the teacher network does not significantly augment the student's learning process. Another hypothesis is the potential inadequacy of the teacher network in providing substantial and meaningful guidance to the student network. The efficacy of knowledge distillation relies on the premise that the teacher network possesses superior knowledge that can be transferred to the student. However, if the teacher network's learned features and representations are not markedly more advanced than those of the student network, the information gap between the teacher and the student remains minimal, implying that the teacher's guidance does not offer novel insights that could enhance the student's performance beyond its current capabilities. The explanation for the observation requires further investigation.

\section{Conclusion and future work}
\label{sec:conclusions}
In the present paper is proposed a method for knowledge distillation (KD) of speech denoising models, based on deep neural networks (DNNs). The proposed method takes advantage of the denoising auto-encoder (DAE) framework, the linear bottlenecks, and the properties of the cosine similarity, and implements a KD process which does not bound the learning of the student to the learned distribution of the teacher and is more robust to numerical scaling of the losses. Additionally, the proposed method works with any dimensionality mismatch between the teacher and the student, which is not true for the SOTA KD methods for speech denoising. 

Further improvements of this method can focus more on the following limitations. Firstly, in this study, the linear bottleneck and the student network are trained concurrently during the KD process, creating conflicting objectives. The linear bottleneck's primary role is to transform the teacher network's latent features, guided by the KD loss. However, this can lead to suboptimal extraction of meaningful information from the teacher's features, as the optimization may prioritize minimizing the KD loss over preserving structured relationships within the teacher's representations. This can result in the student network receiving less valuable information, compromising the KD efficacy. Additionally, with the bottleneck trained along the KD process and cosine distance, which is scale invariant, as the KD loss, there are no constraints on the scale of the transformed teacher features, potentially leading to numerical imbalance.

Using cosine distance as the similarity metric in KD aligns the learned information between the teacher and student in terms of orientation, not feature scale. This allows the student network to learn from the teacher's features without being constrained by their scale, providing greater flexibility and potentially leading to more robust and generalized representations. However, \cite{sahoo:2025:arxiv} shows that cosine similarity may not accurately measure similarity if feature dimensions are correlated. Therefore, while cosine distance has benefits, it is important to recognize its limitations with correlated features. Exploring alternative metrics that account for feature correlations could improve the efficacy of KD.

This study demonstrates the efficacy of employing the linear bottlenecks together with cosine similarity on DAE structure to perform KD in speech denoising. Results show that the proposed method has an improvement, even though marginal, compared to SOTA KD methods. The limitations of the proposed method are acknowledged and will be addressed in the future work.

\bibliographystyle{unsrt}  
\bibliography{references}

\end{document}